%% file: ms.tex
\documentclass[11pt,preprint]{aastex}
\begin{document}
\title{Spectroscopic Studies of Very Metal-Poor Stars with the
Subaru High Dispersion Spectrograph. III. Light Neutron-Capture
Elements\footnote{Based on data collected at the Subaru Telescope,
which is operated by the National Astronomical Observatory of Japan.}}

\author{Wako Aoki\altaffilmark{2}, Satoshi Honda\altaffilmark{2},
Timothy C. Beers\altaffilmark{3},Toshitaka Kajino\altaffilmark{2},
Hiroyasu Ando\altaffilmark{2}, John E. Norris\altaffilmark{4}, Sean
G. Ryan\altaffilmark{5}, Hideyuki Izumiura\altaffilmark{6}, Kozo
Sadakane\altaffilmark{7}, Masahide Takada-Hidai\altaffilmark{8}}

\altaffiltext{2}{National Astronomical Observatory, Mitaka, Tokyo,
181-8588, Japan; email: aoki.wako@nao.ac.jp, honda@optik.mtk.nao.ac.jp,
kajino@th.nao.ac.jp, ando@optik.mtk.nao.ac.jp}
\altaffiltext{3}{Department of Physics and Astronomy, Michigan State
University, East Lansing, MI 48824-1116; email: beers@pa.msu.edu}
\altaffiltext{4}{Research School of Astronomy and Astrophysics, The
Australian National University, Mount Stromlo Observatory, Cotter
Road, Weston, ACT 2611, Australia; email: jen@mso.anu.edu.au}
\altaffiltext{5}{Department of Physics and Astronomy, The Open
University, Walton Hall, Milton Keynes, MK7 6AA, UK; email:
s.g.ryan@open.ac.uk} 
\altaffiltext{6}{Okayama Astrophysical Observatory, National
Astronomical Observatory, Kamogata-cho, Okayama, 719-0232, Japan;
email: izumiura@oao.nao.ac.jp}
\altaffiltext{7}{Astronomical Institute, Osaka Kyoiku University,
Kashiwara, Osaka, 582-8582, Japan; email:
sadakane@cc.osaka-kyoiku.ac.jp}
\altaffiltext{8}{Liberal Arts Education Center, Tokai University,
Hiratsuka, Kanagawa, 259-1292, Japan; email:
hidai@apus.rh.u-tokai.ac,jp}

\begin{abstract} 

Elemental abundance measurements have been obtained for a sample of 18
very metal-poor stars using spectra obtained with the Subaru Telescope
High Dispersion Spectrograph. Seventeen stars, among which 16 are
newly analyzed in the present work, were selected from candidate
metal-poor stars identified in the HK survey of Beers and
colleagues. The metallicity range covered by our sample is $-3.1
\lesssim$ [Fe/H] $\lesssim -2.4$. The abundances of carbon,
$\alpha$-elements, and iron-peak elements determined for these stars
confirm the trends found by previous work. One exception is the large
over-abundance of Mg, Al and Sc found in BS~16934--002, a giant with
[Fe/H] = $-2.8$. Interestingly, this is the most metal-rich star (by
about 1 dex in [Fe/H]) known with such large overabundances in these
elements. Furthermore, BS~16934--002 does {\it not} share the large
over-abundances of carbon that are associated with the two other,
otherwise similar, extremely metal-poor stars CS~22949--037 and
CS~29498--043.

By combining our new results with those of previous studies, we investigate the
distribution of neutron-capture elements in very metal-poor stars, focusing on
the production of the light neutron-capture elements (e.g., Sr, Y, and Zr).
Large scatter is found in the abundance ratios between the light and heavy
neutron-capture elements (e.g., Sr/Ba, Y/Eu) for stars with low abundances of
heavy neutron-capture elements. Most of these stars have extremely low
metallicity ([Fe/H]$\lesssim -3$). By contrast, the observed scatter in these
ratios is much smaller in stars with excesses of heavy neutron-capture elements,
and with higher metallicity. These results can be naturally explained by
assuming that two processes independently enriched the neutron-capture elements
in the early Galaxy. One process increases both light and heavy neutron-capture
elements, and affects stars with [Fe/H]$\gtrsim -3$, while the other process
contributes only to the light neutron-capture elements, and affects most stars
with [Fe/H]$\gtrsim -3.5$. Interestingly, the Y/Zr ratio is similar in stars
with high and low abundances of heavy neutron-capture elements. These results
provide constraints on modeling of neutron-capture processes, in particular
those responsible for the nucleosynthesis of light neutron-capture elements at
very low metallicity.

\end{abstract} 

\keywords{
nuclear reactions, nucleosynthesis, abundances --- stars: abundances
--- stars: Population II}

\section{Introduction}\label{sec:intro}

Very metal-poor stars (hereafter, VMP stars, with [Fe/H] $\le
-2.0$)\footnote{[A/B] = $\log(N_{\rm A}/N_{\rm B})- \log(N_{\rm
A}/N_{\rm B})_{\odot}$, and $\log \epsilon_{\rm A} = \log(N_{\rm
A}/N_{\rm H})+12$ for elements A and B.} belonging to the halo
population of the Galaxy are believed to have formed at the earliest epoch of
star formation, and preserve at their surfaces the chemical
composition produced by the first generations of stars. Studies of the
chemical composition of VMP stars have, in the past decade, proven to
be very important for understanding individual nucleosynthesis
processes \citep[e.g., ][]{mcwilliam95,ryan96,cayrel04}. In
particular, the abundances of the neutron-capture elements provide
strong constraints on the modeling of explosive nucleosynthesis, and
for identifying the likely astrophysical sites in which they are
produced.

One surprising result found by previous studies is the existence of a large
scatter in measured abundance ratios between the neutron-capture elements and
other metals (e.g., Ba/Fe). The scatter appears most significant at [Fe/H]$\sim
-3$. For instance, the abundance ratio of [Ba/Fe] in stars near this
metallicity exhibits a range of about three dex \citep[e.g., ][]{ mcwilliam98}.
Some of the Ba-enhanced stars have abundance patterns that can be explained by
the slow neutron-capture process (s-process). These stars typically also have
high abundances of carbon. Such carbon-enhanced, metal-poor, s-process-rich
(hereafter, CEMP-s) stars are believed to belong to binary systems, the
presently observed member having been polluted by an Asymptotic Giant Branch (AGB)
companion through mass transfer at an earlier epoch. However, even after
removing the CEMP-s stars from samples of objects with metallicity near [Fe/H] =
$-3.0$, a large scatter in [Ba/Fe] remains. 

Even though the abundance ratios between neutron-capture elements and
lighter metals like iron exhibit significant scatter, the abundance
patterns of heavy neutron-capture elements (Ba--Os) in stars with
[Fe/H]$< -2.5$ agree very well with that of the r-process component in
solar system material \citep[e.g.,
][]{sneden96,westin00,cayrel01}. This fact indicates that the
neutron-capture elements in metal-poor, non CEMP-s stars have
originated from the r-process. From the abundance ratios of Ba/Eu or
La/Eu, \citet{burris00} and \cite{simmerer04} concluded that
significant contributions from the (main) s-process appear at
[Fe/H]$\sim -2.3$, though small effect of the s-process is also
suggested at slightly lower [Fe/H] ($\sim -2.7$).

The large scatter in [Ba/Fe] in stars with [Fe/H] $< -2.5$ means that
the astrophysical site responsible for this process was different from
the site responsible for Fe production, and the elements produced were
not well mixed into the interstellar matter from which the low-mass
stars we are currently observing were formed. Moreover, the above
observational result suggests that the abundance pattern of the
solar-system r-process component is not the result of a mixture of
yields from individual nucleosynthesis events having very different
abundance patterns, but rather is a result of nucleosynthesis
processes that yielded very similar abundance patterns for these
elements throughout Galactic history. This phenomenon is sometimes
referred to as the ``universality'' of the r-process nucleosynthesis,
and it is believed to be a key for understanding this process, as well
as for the use of the heaviest radioactive nuclei (Th and U) as
cosmo-chronometers to place direct limits on the ages of extremely
metal-poor (EMP; [Fe/H] $\lesssim -3.0$) and VMP stars.

%Moreover, the abundance patterns of neutron-capture elements for non
%CEMP-s stars at this metallicity are quite similar to that of the
%rapid neutron-capture process (r-process) component in solar system
%material. 
% These facts suggest that the r-process started to operate at
%[Fe/H]$\sim -3$, and 

By way of contrast, measured abundance ratios between {\it light} neutron-capture
elements, such as Sr ($Z=38$), and {\it heavy} ones ($Z\geq 56$) in VMP
stars exhibit a large dispersion \citep[e.g., ][]{mcwilliam98}. The abundance patterns
from Sr to Ba in stars with enhancements of r-process elements do {\it not} agree well
with that of the solar-system r-process component \citep{sneden00}. In other
words, the universality of the r-process apparently does not 
apply to the lighter neutron-capture elements.

The processes responsible for the synthesis of light neutron-capture elements
have been recently studied with growing interest. \citet{truran02} pointed out
the existence of a group of stars that exhibit high Sr/Ba ratios but low Ba
abundances, and emphasized the contrast to r-process-enhanced stars that have
relatively low Sr/Ba ratios. These authors suggested the existence of a quite
different origin of these elements than those (heavier) elements formed by the
``main'' r-process. \citet{travaglio04} investigated the production of light
neutron-capture elements, Sr, Y, and Zr, from the point of view of Galactic
chemical evolution. They concluded that there must exist a process that has
provided these light neutron-capture elements throughout Galactic history, which
they referred to as the ``lighter element primary process (LEPP)'', and
estimated its contribution to the abundances of Sr, Y, and Zr in solar-system
material.

Thus, although the existence of a process that produces light neutron-capture
elements without significant contribution to the heavier ones is suggested by
several previous studies, its astrophysical site remains unclear. This process
is usually distinguished from the weak s-process, which is believed to occur in
helium-burning massive stars, but thought to be inefficient at low metallicity.
It is clearly desirable to obtain additional observational constraints on the
nature of this unknown process for the production of light neutron-capture
elements, such as the elemental abundance patterns produced by the process, and
the level of its contribution to stars with different metallicity. Although some
observational studies \citep[e.g., ][]{johnson02b} have previously focused on
this issue, and have provided important abundance results, more studies for
larger sample of stars with very low metallicity and accurately measured
abundances are required.

Our previous study determined chemical abundances for 22 VMP stars, and
discussed the abundance ratios of neutron-capture elements (Honda et al. 2004a,
b: hereafter, Paper I and Paper II, respectively). In Paper II, we investigated
the correlation between Sr and Ba abundances for stars with [Fe/H]$<-2.5$, and
found that the scatter of the Sr abundance ($\log\epsilon$(Sr)) increases with
decreasing Ba abundance. The present paper reports the chemical abundances for
an additional 16 VMP stars, as well as for two stars that have already been
studied in Paper II, based on observations obtained with the Subaru Telescope
High Dispersion Spectrograph \citep[HDS; ][]{noguchi02} during its commissioning
phase.

In Section~\ref{sec:obs}, we describe the sample selection, details of the
observations, and measurements of equivalent widths and radial velocities.
Elemental abundance measurements are presented in Section~\ref{sec:ana}. In this
section we also consider a new star that exhibits a significantly high Mg/Fe
abundance ratio compared to other VMP stars in our study. In
Section~\ref{sec:disc} we combine the chemical abundances for VMP stars
reported by previous work with our new sample, and discuss the production of light
neutron-capture elements in the early Galaxy.

\section{Observations and Measurements}\label{sec:obs}

\subsection{Sample Selection and Photometry Data}

Our sample of stars was originally selected from candidate very metal-poor stars
identified in the HK survey of Beers and colleagues (Beers, Preston, \& Shectman
1985; 1992; Beers 1999) whose medium-resolution (1--2~{\AA}) spectra indicated
that they possessed metallicities [Fe/H] $\le -2.5$. While in Paper II we
selected objects that were likely to have excesses of r-process elements, our
new sample has no such explicit bias. Indeed, none of the objects in the new
sample exhibit significant excesses of heavy neutron-capture elements (see
below). Nevertheless, the abundances of light neutron-capture elements such as Sr
are distributed over a very wide range in our new sample.  Table 1 provides a
listing of the objects considered in our study, including their coordinates,
details of the observations conducted, and their measured radial velocities (see
below).

The neutron-capture element-enhanced star CS~30306--132 was already analyzed in
Paper II, but is also included here for comparison purposes. For the same
reason, the bright metal-poor giant HD~122563 was also analyzed.

Table 2 presents optical $BVRI$ photometry (Johnson--Kron--Cousins system)
for our sample stars; with the exception of HD~122563, these data are
drawn from photometry reported by Beers et al. (2005, in
preparation). Errors in the $BVRI$ magnitudes are typically on the
order of 0.01--0.02 magnitudes. Near infrared $JHK$ photometry was,
again with the exception of HD~122563, provided by the Two Micron All
Sky Survey (2MASS) Point Source Catalog \citep{skrutskie97}. Estimates
of the interstellar reddening, $E(B-V)$, for each object were obtained
from the \citet{schlegel98} map; the extinction of each band was
obtained from the reddening relation provided in their Table 6.

\subsection{High-Dispersion Spectroscopy}

High-dispersion spectroscopy for the purpose of conducting our chemical
abundance studies was obtained with Subaru/HDS, using a spectral resolving power
$R = 50,000$ (a slit width of 0.72~arcsec), in April 2001, July 2001, and February
2002. The atmospheric dispersion corrector (ADC) was used in all observing runs.
Two EEV-CCDs were used with no on-chip binning. The spectra cover the wavelength
range 3550--5250 {\AA}, with a small gap in the coverage between 4350 and
4450~{\AA} due to the physical separation between the two CCDs.

The object list and observing details are given in Table 1. Standard
data reduction procedures (bias subtraction, flat-fielding, background
subtraction, extraction, and wavelength calibration) were carried out
with the IRAF echelle package\footnote{IRAF is distributed by the
National Optical Astronomy Observatories, which is operated by the
Association of Universities for Research in Astronomy, Inc. under
cooperative agreement with the National Science Foundation.}. In order
to remove suspected cosmic-ray hits, we first apply median filtering
to a two-dimensional CCD image. When a remarkably high count was found
at one pixel in the original image compared to the median-filtered
image, the recorded counts of that pixel were replaced by the value
obtained in the median-filtered image.\footnote{At the beginning of
the data reduction, we subtracted bias level from each object
frame. Since the inter-order region of the object frame has very low
count, we first added a constant to object frame, and then applied the
IRAF task {\it median} to make median-filtered image (with the
parameter {\it x[y]window} of 3). The constant added to the object
frame is the same order of the peak count due to the photons from the
star. We divided the object frame by the median-filtered image,
resulting in a frame which has count close to unity, with exceptions
of pixels affected by cosmic-ray. We identified the pixels that have
counts more then 20~\% higher than those in surrounding pixels as
those affected by cosmic-ray, and replaced their counts by unity
(this process is more effectively made by applying the task {\it
lineclean} to fitting to both column and line directions). We multiply
the obtained frame by the median-filtered image, and finally subtract
the constant that was added at the first process. We confirmed the
above parameter choices are quite conservative, i.e., this procedure does not
affect the pixels which are apparently not affected by cosmic-ray
events, though there remain some pixels which seem to be affected by
cosmic-ray.} Wavelength calibration was performed using Th-Ar spectra
obtained a few times during each night of observations.

The signal-to-noise (S/N) ratio given in Table~\ref{tab:obs} was estimated from
the peak counts of the spectra in the 149th Echelle order ($\sim 4000$~{\AA}).
We note that the values are given per resolution element (6~km~s$^{-1}$).
Since the resolution element is covered by about 6.7 pixels, the S/N ratios per
pixel are by a factor of $\sim 2.6$ lower than those listed in the table.

\subsection{Equivalent Widths}\label{sec:ew}

Equivalent widths were measured for isolated atomic lines by fitting
gaussian profiles \citep{press96}, while a spectrum synthesis technique
was applied to CH molecular bands, as well as to atomic lines that are
significantly affected by hyperfine splitting. The measured equivalent
widths of elements lighter than La ($Z\leq 56$) are given in
Table~\ref{tab:ew}. Heavier elements are detected only in stars having
relatively higher abundances of neutron-capture elements. The measured
equivalent widths of heavier elements are given for 11 objects in
Table~\ref{tab:ew2} separately.

Comparisons of equivalent widths with those reported in Paper I are
shown in Figure 1 for HD~122563 and CS~30306--132. While the same data
were analyzed for CS~30306--132 in both studies, our data for
HD~122563 are different than those in Paper I. Measurements of
equivalent widths were made independently by W.A. (this work) and
S.H. (Paper I) using different software, although both applied
gaussian fitting procedures. The two measurements show quite good
agreement, although small departures appear for the strongest lines;
the trends are opposite for the two stars, suggesting these are not
systematic in origin.  Most likely, they are the result of adopting
slightly different fitting ranges for strong lines by the two
investigators.

%local continuum
%estimates 

Comparisons with the measurements by \citet{cayrel04} are also shown in Figure 2
for HD~122563 and CS~30325--094. The agreement is again quite good; there is no
obvious systematic differences between the two sets of measurements. 

The equivalent widths of the two resonance lines of Ba require special
attention. Comparisons of the equivalent widths between the two lines
are shown in Figure~\ref{fig:ewba}. The equivalent width of the
\ion{Ba}{2} 4934~{\AA} line is sometimes larger than that of the
4554~{\AA} line, even though the $gf$-value of the 4934~{\AA} is
smaller, by a factor of two, than that of the 4554~{\AA} line. This
apparent discrepancy is not necessarily due to measurement
error. First, the wavelength of the former line is larger by $\sim$8\%
than the latter one, and correction for this factor is required in
order to compare the equivalent widths. In addition, the effect of
hyperfine splitting is expected to be more significant for the
4934~{\AA} line than that for the 4554~{\AA} line, if isotopes with
odd mass number ($^{135}$Ba and $^{137}$Ba) significantly contribute
to the absorption. We simulated these effects by calculating
equivalent widths of both lines using a model atmosphere for a
metal-poor giant, assuming Ba isotope fractions of the r-process
component in solar-system material \citep{arlandini99}. The results
are shown by the solid line in Figure~\ref{fig:ewba}. Our calculations
demonstrate that, after the small correction for the difference of the
wavelengths, the equivalent width of the 4554~{\AA} is twice that of
the 4934~{\AA} line in the weak-line limit ($W\lesssim 20$~m{\AA}),
but the equivalent widths of the two lines are similar at $W\sim
120$~m{\AA}; the equivalent width of the 4934~{\AA} is larger than
that of 4554~{\AA} if the lines are stronger. We also show the results
of calculations not including hyperfine splitting for comparison
purposes (the dashed line in the figure). The measured equivalent
widths of strongest lines are better explained by the calculations
taking account of the effect of hyperfine splitting. This suggests a
large contribution from isotopes with odd mass numbers, which are
expected as a result of r-process nucleosynthesis. This would be
consistent with the fact that the objects having strong Ba lines in
our sample show abundance ratios (e.g., Ba/Eu ratios) similar to that
of the r-process component in solar system material (see below).

%as found in Table~\ref{tab:ew} (e.g., column 10 = CS~22886--042)

%Our measurements of Ba equivalent widths in stars with strong Ba lines

\subsection{Radial velocities}

We measured heliocentric radial velocities ($V_{\rm r}$) for each spectrum, as
given in Table~\ref{tab:obs}. The measurements were made using clean, isolated
\ion{Fe}{1} lines. The standard deviation of the values from individual lines is
adopted as the error of the measurements reported in the table. Systematic
errors in the measurements are not included in these errors. The wavelength
calibration was made using Th-Ar comparison spectra that were obtained during
the observing runs, without changing the spectrograph setup. Hence, the
systematic error is basically determined by the stability of the spectrograph.
The spectrum shift during one night is at most 0.5 pixel (0.45~km~s$^{-1}$),
which corresponds to a temperature variation of four degrees centigrade
\citep{noguchi02}, if the setup is not changed during the night. Combining this possible
systematic error with the random errors (the 3~$\sigma$ level is typically
0.6-0.8~km~s$^{-1}$), the uncertainties of the reported radial velocity 
measurements are $0.7\sim 1.0$~km~s$^{-1}$.

For three stars in our sample (BS~16934--002, CS~30306--132, and CS~30325--028),
two or three spectra were obtained on different observing runs. While no clear
variation of $V_{\rm r}$ is found in BS~16934--002 and CS~30306--132,
CS~30325--028 exhibits a 2.4~km~s$^{-1}$ difference between the two
measurements, suggesting possible binarity of this object. Further monitoring of
radial velocity for this object is required to determine its binary nature,
which may be related to its chemical abundance properties.

The heliocentric radial velocity of HD~122563 measured in our study is $V_{\rm
r}=-25.81$~km~s$^{-1}$. This is similar to the previous measurements reported in
Paper I and in \citet{norris96} ($V_{\rm r}=-26.5 \sim -27.2$~km~s$^{-1}$).
A radial velocity $V_{\rm r}=-108.46$~km~s$^{-1}$ was obtained for CS~30306--132 from the 2001
July spectrum. This agrees, within the errors, with the results obtained from
the independent measurement reported in Paper I using the same
spectra.\footnote{Note that the sign of the $V_{\rm r}$ of CS~30306--132 in Paper
I is not correct: the correct value is $V_{\rm r}=-109.01$~km~s$^{-1}$.}

\section{Chemical Abundance Analysis and Results}\label{sec:ana}

A standard analysis using model atmospheres was performed for the
measured equivalent widths for most of the detectable elements, while
a spectrum synthesis technique was applied for the CH molecular bands
and atomic lines strongly affected by hyperfine splitting. For the
calculation of synthetic spectra and equivalent widths using model
atmospheres, we applied the LTE spectral synthesis code used in
\citet{aoki97}. \citet{unsold55}'s treatment of van der Waals
broadening, enhanced by a factor of 2.2 in $\gamma$, was used as in
\citet{ryan96}. The polynomial partition function approximations provided by
\citet{irwin81} were applied to the heavy elements. In this section we
describe the determination of stellar atmospheric parameters
(subsection~\ref{sec:param}) and abundance measurements for carbon
(subsection~\ref{sec:carbon}), $\alpha$-elements
(subsection~\ref{sec:alpha}), and the neutron-capture elements
(subsection~\ref{sec:ncap}) in detail. Estimates of uncertainties in
abundance measurements are presented in subsection~\ref{sec:error}.

%The results are listed in Table~\ref{tab:res} and \ref{tab:heavy}.

\subsection{Atmospheric Parameters}\label{sec:param}

Effective temperatures were estimated from the photometry listed in
Table~\ref{tab:photometry} using the empirical temperature scale of
\citet{alonso99}, after reddening corrections were carried out. The
filter-corrections of \citet{fernie83} were applied to convert the
Johnson-Kron-Cousins $V-R$, $V-I$, and $R-I$ colors to Johnson ones
that were used in the temperature scale of \citet{alonso99}. The
filter-corrections provided by \citet{carpenter01} and
\citet{alonso94} were applied to 2MASS $J, H$, and $K$ data to derive
those in the TSC system (via the CIT ones) that is used by
\citet{alonso99}. We have chosen to adopt the values
determined from $V-K$, as described in Paper II
(Table~\ref{tab:photometry}). For comparison purposes, we also give
the difference between the effective temperature from the $V-K$ and
the average of the effective temperature from four colors ($B-V, V-R,
V-I$ and $R-I$). The agreement is quite good, less than $\pm 100$ K,
between the effective temperatures derived from $V-K$ and from the
average of the optical bands.

The effective temperatures adopted in the abundance analyses are listed in
Table~\ref{tab:param}. Note that the near infrared photometry data were not
available for several objects when our first abundance analyses were made. In
these cases, we estimated the effective temperature from the optical colors
(e.g., $B-V$), and performed a re-analysis if the effective temperature obtained
from $V-K$ is significantly different from that adopted in our first analyses.
In other cases, we did not repeat the analysis. The largest difference of the
effective temperature from $V-K$ and adopted one is 79~K (CS~30306--132), which
is below the expected error of the effective temperature determination.

An LTE abundance analysis was carried out for \ion{Fe}{1} and
\ion{Fe}{2} lines using the model atmospheres of \citet{kurucz93}.
We performed abundance analyses in the standard manner for the measured
equivalent widths. Surface gravities ($\log g$) were determined from the
ionization balance between Fe {\small I} and Fe {\small II}; the microturbulent
velocity ($v_{\rm tur}$) was determined from the Fe {\small I} lines by
demanding no dependence of the derived abundance on equivalent widths. The final
atmospheric parameters are reported in Table~\ref{tab:param}.

We note that there exists a correlation between the lower excitation
potential ($\chi$) of \ion{Fe}{1} lines and the abundance derived from
individual lines. The slope is typically $-0.04$~dex~eV$^{-1}$. Such
correlations were already reported by \citet{johnson02a} who also
applied the Kurucz's model atmosphere grid to the analyses of very
metal-poor giants. This trend disappears if systematically lower
effective temperatures (by about 150~K) are assumed
\citep{johnson02a}. Hence, our effective temperatures might be
systematically higher than those determined spectroscopically in
previous studies.

\subsection{Carbon}\label{sec:carbon}

Carbon abundances were estimated from the CH molecular band at 4322~{\AA}
following the procedures described in Paper II, using the same line list 
for the CH band. The band was detected in all stars except for CS~30327--038,
for which only an upper limit of the carbon abundance could be estimated. 

The carbon abundances of HD~122563 and CS~30306--132 were also
determined by Paper II. The present work adopts similar atmospheric
parameters to those in the previous one, and the agreement between the
two measurements is fairly good.

The carbon abundances of HD~122563 and CS~30325--094 were also
measured by \citet{cayrel04} from the G-band of the CH $A-X$
system. While the agreement of the [C/Fe] values for HD~122563 between
the two works is fairly good, the [C/Fe] determined by our present
analysis for CS~30325--094 is 0.5~dex higher than that of
\citet{cayrel04}. The discrepancy can be partially ($\sim$0.2~dex)
explained by the small difference of the adopted effective
temperatures (100~K). However, the reason for the remaining
discrepancy is not clear.

The carbon abundances of giants are expected to be affected by
internal processes, i.e., CNO-cycle and dredge-up.
Figure~\ref{fig:cl} shows the correlation between the carbon abundance
ratio ([C/Fe]) and luminosity ($\log L$/L$_{\odot}$) that is estimated
using the relation $L/L_{\odot}\propto (R/R_{\odot})^{2}(T_{\rm
eff}/T_{{\rm eff}\odot})^{4}\propto
(M/M_{\odot})(g/g_{\odot})^{-1}(T_{\rm eff}/T_{{\rm eff}\odot})^{4}$,
assuming the mass of the stars to be 0.8~M$_{\odot}$, for our sample
and those of Paper II and \citet{cayrel04}. While the bulk of stars
with $\log L$/L$_{\odot} \lesssim 2.5$ have [C/Fe]$\sim +0.4$, [C/Fe]
decreases with increasing luminosity.\footnote{A few stars have
exceptionally high carbon abundances ([C/Fe]$\sim +1.0$) at high
luminosity ($\log L$/L$_{\odot} \sim 2.7$). The well known carbon-
enhanced objects CS~22892--052 and CS~22949--037 are included in this
group.} This decreasing trend can be interpreted as a result of
internal processes. Similar tendency was already found by
\citet{cayrel04} who investigated the correlation between [C/Fe] and
effective temperature for their sample. However, the trend is more
clear in our figure where luminosity is adopted as the indicator of
the evolutionary stage. A more detailed analysis for the Cayrel et
al.'s sample was made by \citet{spite05}, including N and Li
abundances. A similar discussion is also found in \citet{carretta00}
for more metal-rich stars.

%from the effective temperature and gravity

\subsection{The $\alpha$ and Iron-Peak Elements}\label{sec:alpha}

There are numerous previous studies of the $\alpha$- and iron-peak elements in
VMP stars. Our measurements confirm the usual over-abundances of
$\alpha$--elements relative to iron that have been found previously for the
majority of metal-poor stars. Figure~\ref{fig:mgfe} shows the trend of [Mg/Fe]
as a function of [Fe/H]. A Mg over-abundance of about 0.4--0.6~dex is found for
most stars in our sample. These values are similar to the majority of metal-poor
dwarf stars reported by \citet{cohen04} and to the giants studied in Paper II.
Note that these two studies found several stars with low abundances of
$\alpha$-elements ([Mg/Fe]$\sim0.0$), while no such star is found in the present
sample.

The [Mg/Fe] values of giants studied by \citet{cayrel04} seem to be slightly
(0.10--0.15~dex) lower than those of our stars. Since only two stars are in
common between the two studies, this discrepancy may be simply due to
differences in the samples under consideration. However, the [Mg/Fe] of
HD~122563 determined by us is +0.54, while \citet{cayrel04} derived
[Mg/Fe]=+0.36 for the same object ($\Delta$ [Mg/Fe]=0.18~dex). Hence, there
seems to exist a systematic difference between the two studies. We note that the
[Mg/Fe] of HD~122563 determined in Paper II ([Mg/Fe]=+0.54) agrees with our
result.

Measurements of Mg/Fe ratios are relatively insensitive to the adopted
atmospheric parameters (subsection~\ref{sec:error}). Hence, the
difference of model parameters between our study and \citet{cayrel04}
for metal-poor giants is not likely to be the source for the
discrepancies in the derived [Mg/Fe] values. However, it should be
noted that the Mg lines used in the abundance measurements in these
two studies are somewhat different. For instance, the present study
uses the \ion{Mg}{1} 4057.5 and 4703.0~{\AA} lines, which were not
adopted by \citet{cayrel04}, while they used \ion{Mg}{1} 4351.9 and
5528.405~{\AA} lines that are not covered by our spectral
range. Moreover, the $gf$-values of some lines are different (e.g.,
\ion{Mg}{1} 4570~{\AA} line: their $\log gf$ value is 0.3~dex higher
than ours).  These differences may partially explain the discrepancies
in the Mg/Fe ratios. We attempted an analysis of Mg abundance for
HD~122563 using the equivalent widths and line data of
\citet{cayrel04}, and derived [Mg/Fe]=+0.45. The discrepancy ($\Delta$
[Mg/Fe]=0.09~dex) then becomes much smaller than the original one.
Nevertheless, there seems to remain a $\sim$0.1~dex discrepancy in
[Mg/Fe] between the two studies for which no clear reason has been
found.

We call attention to the very large Mg over-abundance in BS~16934--002
([Mg/Fe]=+1.25).  Figure~\ref{fig:sp_mg} shows examples of Mg
absorption features in this star, compared to those of HD~122563,
which has similar atmospheric parameters and iron abundance. While the
strengths of the Fe absorption lines are very similar in the two
objects, the Mg absorption lines in BS~16934--002 are clearly much
stronger than those in HD~122563. We note that the lower excitation
potential of the two Mg lines shown in this figure are quite
different, hence the stronger Mg absorption features in BS~16934--002
is not due to the (small) differences of atmospheric parameters in the
two stars. The derived Ti abundance of BS~16934--002 relative to Fe
is slightly higher than other stars in our sample, including
HD~122563. This is also seen in the spectra shown in
Figure~\ref{fig:sp_mg}.  Another remarkable abundance anomaly found in
BS~16934--002, compared to other stars, is its high abundances of Al
([Al/Fe]=+0.03) and Sc ([Sc/Fe]=+0.7).

% <-- WHY DO WE NOT SHOW THE SAME WAVELENGTH RANGES IN THIS FIGURE ?

There are two other well-studied stars with [Fe/H]=$-3.5 \sim -4.0$ having
extremely high Mg/Fe ratios (CS~22949--037: McWilliam et al. 1995, Norris et al.
2001, Depagne et al. 2002; CS~29498--043: Aoki et al. 2002a). These two stars
also exhibit large over-abundances of C, N, O, and Si, hence they might be more
properly interpreted as ``iron-deficient'' stars, perhaps related to supernovae
that ejected only small amount of material from their deepest layers
\citep{tsujimoto03,umeda03}. It is thus of some significance that BS~16934--002
has {\it no clear excess} of either C or Si. Its iron abundance is more than
five times higher than the two stars. Hence, the origin of the Mg excess in
BS~16934--002 is perhaps quite different from that in CS~22949--037 and
CS~29498--043. Further chemical abundance measurements based on higher quality
spectra are clearly needed to understand the nucleosynthesis processes
responsible for this star. In particular, oxygen would be a key element to
measure.

\subsection{The Neutron-Capture Elements}\label{sec:ncap}

The abundances of Sr and Ba were determined by a standard analysis of
the \ion{Sr}{2} and \ion{Ba}{2} resonance lines. Because of their
large transition probabilities and the relatively high abundances of
these two elements in our stars, Sr and Ba are detected in all of our
program objects. An exception is the Ba in CS~30325--094, for which
only an upper limit of its abundance was determined. In the analyses
of the Ba lines, the effects of hyperfine splitting and isotope shifts
are included, assuming the isotope ratios of the r-process component
of solar system material. The Ba in such VMP stars is expected to have
originated from the r-process, except for stars with large excesses of
carbon and s-process elements \citep[e.g., ][]{mcwilliam98}. Since the
metallicity range covered by our sample is [Fe/H]$\leq-2.4$, where
contributions of the s-process is small in general (see Section 1),
and no CEMP-s stars are included in our sample, the above assumption
is quite reasonable for our analyses. The effect of hyperfine
splitting is clearly seen in stars with strong Ba lines, as mentioned
in subsection~\ref{sec:ew}.

The light neutron-capture elements Y and Zr are detected in 14 and 11 stars in
our sample, respectively. The derived abundances of these four elements (Sr,
Y, Zr, and Ba) are listed in Table~\ref{tab:res}. An upper-limit of the
abundance is given when no absorption line is detected.

Our sample includes no stars with significant over-abundances of heavy
neutron-capture elements, with the exception of CS~30306--132, which
was already studied in Paper II and is re-analyzed here for comparison
purposes. For this reason, elements heavier than Ba are detected only
in a limited number of objects. The abundances of these heavy elements
are given in Table~\ref{tab:heavy}. The abundances were determined by
standard analysis techniques, taking into account hyperfine
splitting for La \citep{lawler01a} and Yb (Sneden 2003, private
communication). For the element Eu a spectrum synthesis technique was
applied because the three \ion{Eu}{2} lines analyzed in the present
work show remarkably strong effects of hyperfine splitting
\citep{lawler01b}. An isotope ratio of $^{151}$Eu:$^{153}$Eu=50:50 was
assumed in the analysis.

Figure~\ref{fig:abpat1} shows the abundance patterns of seven elements from Zn
to Eu for BS~16543--097 and BS~16080--054, whose Zn abundances are quite
similar. BS~16543--097 is a star having relatively high abundances of heavy
neutron-capture elements as compared to most stars in our sample, while the
abundances of these elements in BS~16080--054 are lower by about 1~dex than
those in BS~16543--097. Nevertheless, the abundance patterns from Zn to Zr are
very similar in the two stars. Such large difference in the abundance ratios
between light and heavy neutron-capture elements has already been reported in a
number of VMP stars \citep[e.g., ][]{truran02}, and suggests the existence of
two (or more) processes that produce neutron-capture elements. This point is
discussed in detail in Section~\ref{sec:disc}.

\subsection{Uncertainties}\label{sec:error}

Random abundance errors in the analysis are estimated from the standard
deviation of the abundances derived from individual lines for each species.
These values are sometimes unrealistically small, however, when only a few lines
are detected. For this reason, we adopted the larger of (a) the value for the
listed species and (b) that for \ion{Fe}{1} as estimates of the random errors. 
Typical random errors are on the order of 0.1~dex.

We estimated the upper limit of the chemical abundances for several elements
when no absorption line is detected. The error of equivalent width measurements
is estimated using the relation $\sigma_{W}\sim (\lambda n_{\rm pix}^{1/2})
/(R[S/N])$, where $R$ is the resolving power, $S/N$ is the signal-to-noise ratio
per pixel, and $n_{\rm pix}^{1/2}$ is the number of pixels over which the
integration is performed \citep{norris01}. The upper limit of equivalent widths,
used to estimate the upper limit of abundances, is assumed to be 3$\sigma_{W}$.
 
Errors arising from uncertainties of the atmospheric parameters were
evaluated for $\sigma (T_{\rm eff})=100$~K, $\sigma (\log g)=0.3$~dex,
and $\sigma (v_{\rm tur}) =0.3$~km s$^{-1}$ for HD~122563,
CS~30306--132, and CS~29516--041.  HD~122563 is a well-known
metal-poor giant. CS~30306--132 has high abundances of neutron-capture
elements, while these elements in CS~29516--041 are relatively
deficient. We applied the errors estimated for elements other than
neutron-capture elements for HD~122563 to all other stars. The
strengths of absorption lines of neutron-capture elements show a quite
large scatter in our sample.  Since the errors in abundance
measurements are sensitive to the line strengths, we estimated errors
in abundance measurements including the difference of line strengths
as follows.  (1) Light neutron-capture elements -- we applied the
errors for Sr, Y, and Zr estimated for HD~122563 to most stars. For
stars with weak Sr lines, we adopted the errors estimated for
CS~29516--041 (in such objects, Y and Zr are not detected). (2) Heavy
neutron-capture elements -- we applied the errors estimated for
CS~30306--132 to stars with strong Ba lines, while we adopted errors
estimated for HD~122563 for other stars.

Finally, we derived the total uncertainty by adding, in quadrature, the
individual errors, and list them in Tables \ref{tab:res} and \ref{tab:heavy}.

\section{Discussion}\label{sec:disc}

In this section we focus on the elemental abundances of light and heavy
neutron-capture elements in metal-poor stars, which are not significantly
affected by the (main) s-process, and discuss their possible origins. We first
inspect the full sample based on the abundances of Sr and Ba, taken to be
representative of the light and heavy neutron-capture elements, respectively,
combining our new measurements with the results of previous work
(subsection~\ref{sec:srba}). Then, we confirm the similarity of the abundance
patterns of light neutron-capture elements in VMP stars with high and low abundances
of heavy neutron-capture elements (subsection~\ref{sec:sryzr}). Since the measured
abundances of Sr and Ba are, unfortunately, rather uncertain, because of the
strengths of the resonance lines, particularly in stars with high abundances of
neutron-capture elements, we investigate the abundances of Y, Zr, and Eu in
detail for stars having relatively high abundances of neutron-capture
elements (subsection~\ref{sec:yzreu}). 

\subsection{Sr and Ba abundances}\label{sec:srba}

As mentioned in Section~\ref{sec:intro}, Sr and Ba abundances in VMP stars
exhibit remarkably large scatter; even the Sr/Ba ratio has a large scatter.
Though the scatter in Sr/Ba ratios appears to be somewhat larger at lower
metallicity, the behavior is unclear, in particular at [Fe/H]$<-3.5$ where the
sample is still very small.

Previous studies \citep[e.g., ][ Paper II]{truran02} have shown that
the Sr/Ba ratio exhibits a correlation with the abundance of Ba (i.e.,
heavy neutron-capture elements), and the scatter is larger at lower Ba
abundance in metal-poor stars. Figure~\ref{fig:srba} shows the
abundances of Sr and Ba for our sample and others from previous
studies, in which stars with [Fe/H]$>-2.5$ are excluded to select only
VMP stars to which contributions of the main s-process are small. The
same diagram was shown in Paper II, but our new sample increases the
number of stars with low Ba abundances. We here adopt the values of
$\log \epsilon$(X), rather than [X/Fe], because the abundances of
neutron-capture elements do not show a clear correlation with Fe
abundance. Moreover, abundances of VMP stars are usually expected to
be determined by a quite limited number of nucleosynthesis events. If
this is true, the abundance ratio relative to Fe is less meaningful,
and indeed makes the discussion more complicated. We discuss the
correlation with metallicity (Fe abundance) later in this section.

As already shown in Paper II, the diagram of Sr and Ba abundances
(Figure~\ref{fig:srba}) clearly demonstrates (1) the absence of
Ba-enhanced stars with low Sr abundance, and (2) the larger scatter in
Sr abundances at lower Ba abundance.\footnote{Some CEMP-s stars
exhibit very high Ba abundances with a moderate excess of Sr (e.g.,
LP~706--7 = CS~31062--012: $\log \epsilon$(Ba)=1.65, and $\log
\epsilon$(Sr)=0.67; Aoki et al. 2002b). These stars are excluded from
our sample, as mentioned in the caption of Figure~\ref{fig:srba}.} The
former result implies that the process that produces heavy
neutron-capture elements like Ba also forms light neutron-capture
elements such as Sr. This gives a strong constraint on the modeling of
the r-process yielding heavy neutron-capture elements, often referred
to as the "main" r-process \citep[e.g., ][]{truran02}.

The distribution of the Sr and Ba abundances in Figure~\ref{fig:srba} can be
naturally explained by assuming two nucleosynthesis processes. One produces both
Sr and Ba, while the other produces Sr with little Ba, as already discussed in
Paper II. In order to investigate this point in more detail, we show Sr-Ba
diagrams separately for three metallicity ranges: [Fe/H]$\leq -3.1$,
$-3.1<$[Fe/H]$\leq -2.9$, and [Fe/H] $> -2.9$ (Figure~{\ref{fig:srba3}). As
can be seen in these figures, the stars with lowest Fe abundances have very low
Ba abundances, while the Ba-rich stars appear at around [Fe/H]$\sim -3$, and
then all stars with [Fe/H]$>-2.9$ have relatively high Ba abundances. This
suggests that the main r-process operates only at [Fe/H] $\gtrsim -3$, and the
effect is more or less seen in all stars with higher metallicity. Similar points
have already been made by previous papers \citep[e.g., ][]{qian00} to
explain the large scatter of Ba abundances in stars at [Fe/H]$\sim -3$. An
important result of the present analysis is that the scatter of Sr abundances
persists even in the lowest metallicity regime. In other words, the presumed
second process that produces Sr with little Ba operates even at [Fe/H]$<-3$ .
This is a clear difference of this process from the main r-process, which
apparently did not significantly affect this metallicity range.

We note that the three stars with [Fe/H]$\sim -4$ studied by
\citet{francois03} have very low abundances of {\it both} Sr and Ba.
The [Fe/H]$\sim -4$ star CS~22949--037 has a relatively high Sr
abundance ($\log \epsilon$(Sr)=$-0.72$, Depagne et al. 2002). However,
the high abundances of C, N, O, and $\alpha$-elements relative to Fe
found in this star mean that Fe is not a good metallicity indicator
for this object. If this object is excluded, all stars with high Sr
abundances in the most metal-poor group (top panel of
Figure~\ref{fig:srba3}) have [Fe/H]$\gtrsim -3.5$. This may be another
constraint on the process producing light neutron-capture elements at
very low metallicity. However, the sample is still too small, and
further measurements of Sr and Ba abundances at [Fe/H]$<-3.5$ are
strongly desired.

The above inspection demonstrates that the process forming both light and heavy
neutron-capture elements (the main r-process) appears only for stars with
[Fe/H]$\gtrsim -3$, while another process producing light neutron-capture
elements appears at even lower metallicity. This metallicity dependence is
important information to constrain the progenitor stars responsible for such
events. However, there exists some controversy on the implication of the
metallicity of these stars. In this metallicity range, no clear age-metallicity
relation can be assumed, since the metal enrichment is expected to be strongly
dependent on the nature of the previous-generation stars and the formation
processes of the low-mass stars we are currently observing.

One possibility is that the metallicity indicates the sequence of mass
of the progenitor stars that contributed to the next generation
low-mass stars.  \citet{tsujimoto00} suggested that lower-metallicity
stars reflect the yields of supernovae whose progenitor mass is lower,
on the basis of the theoretical prediction that supernovae from lower
mass progenitors yield smaller amount of
metals.\footnote{\citet{tsujimoto00} adopted [Mg/H] as a metallicity
indicator, while our discussion makes use of [Fe/H]. However, the
[Mg/Fe] ratio is almost constant in most stars in this metallicity
range, as seen in subsection \ref{sec:alpha}.} The high abundances of
light neutron-capture elements in some stars in the lowest metallicity
range indicate that these elements were provided by supernovae with
even lower-mass progenitors, while the main r-process is related to
progenitors with 20~M$_{\odot}$, according to \citet{tsujimoto00}. On
the other hand, the lower metallicity might result from the higher
explosion energy of type II supernovae, which swept up larger amounts
of interstellar matter and induced the formation of next-generation
stars with lower metallicity. If higher-mass progenitor stars explode
with higher energy, the existence of stars with high abundances of
light neutron-capture elements at very low metallicity indicates that
the process responsible for these elements is related to very massive
stars.

Because of the difficulties noted above, interpretations of the
metallicity dependence of the processes producing light and/or heavy
neutron-capture elements are still premature. Although our results
provide constraints on such models, further investigation of each
process is required. In particular, detailed chemical abundance
studies of stars having high abundances of neutron-capture elements
would be very useful. Previously, several stars with large
overabundances of heavy neutron-capture elements have been studied in
great detail \citep[e.g., ][]{sneden96,cayrel01}, providing strong
constraints on models of the main r-process. By way of contrast,
studies for stars with low Ba and high Sr abundances are still quite
limited to date. Studies of the detailed abundance patterns of such
objects, as has been made for stars with excesses of heavy r-process
elements like CS~22892--052 \citep{sneden03} and CS~31082--001
\citep{hill02}, will provide a definitive constraint on modeling the
presumed additional process that creates the light neutron-capture
elements in the early Galaxy.

\subsection{Abundance Ratios of Light Neutron-Capture Elements}\label{sec:sryzr}

In the previous section we investigated the correlation between the
abundances of Sr and Ba, which are detected in almost all stars in our sample.
In this section we investigate the abundance ratios of the three light
neutron-capture elements Sr, Y, and Zr ([Sr/Y] and [Y/Zr]). 

Figure~\ref{fig:sryzr} shows the values of [Sr/Y] and [Y/Zr] as
functions of [Ba/H] for stars with [Fe/H] $<-2.5$. Since Y and Zr are
detected only in stars with relatively high Sr abundances, stars with
very low Sr abundance, which are located at the lower left in
Figure~\ref{fig:srba}, are not included in
Figure~\ref{fig:sryzr}. Therefore, the stars with high [Ba/H] reflect
the results of the main r-process, while the other process that
produces light neutron-capture elements with little heavier species is
presumed to be responsible for stars with low [Ba/H] values in this
figure. The lower panel of Figure~\ref{fig:sryzr} shows no clear
dependence of [Y/Zr] on [Ba/H], indicating that these two processes
produce very similar abundance ratios of light neutron-capture
elements.  The average value of [Y/Zr] for the 7 stars in our sample
in which both Y and Zr are detected is $<$[Y/Zr]$>=-0.44$, while that
of the all stars shown in the figure (29 stars) including objects
studied by previous work is $-0.38$.

The Y and Zr abundance ratios in VMP stars have been investigated by
\citet{johnson02b} in some detail. These authors found no correlation between
[Y/Zr] and [Fe/H] in the very low metallicity range. This means that there is no
correlation between [Y/Zr] and the abundances of heavy neutron-capture elements,
because their sample includes stars with both high and low abundances of Ba.
Figure~\ref{fig:sryzr} shows this result more clearly, by increasing the sample
of VMP stars and by adopting the Ba abundance, instead of Fe, as the horizontal
axis.

Since the main component of the s-process is the dominant contributor
to Y and Zr in solar-system material \citep[e.g., ][]{arlandini99},
the [Y/Zr] ratio yielded by this process is similar to the
solar-system one ([Y/Zr]$_{\rm main-s} \sim 0$). This is clearly
different from the [Y/Zr] value found in the VMP stars shown in
Figure~\ref{fig:sryzr}. The weak component of the s-process, which was
introduced to explain the excess of light s-process nuclei in the
Solar System, is a candidate for the process responsible for the stars
with large excesses of light neutron-capture elements relative to the
heavier ones. However, the yields of elements produced by this process
rapidly decreases with increasing mass number at around $A\sim 90$.
Indeed, the [Y/Zr] ratio predicted by \citet{raiteri93} for the weak
s-process is [Y/Zr]$_{\rm weak-s}$=+0.3, which cannot explain the
observational results for VMP stars.

The [Y/Zr] ratio predicted for the r-process component in the Solar
System, estimated by \citet{arlandini99}, is [Y/Zr]$_{\rm r-ss}\sim
-0.3$, which agrees well with the values found for VMP stars. This
suggests that the light neutron-capture elements in VMP stars,
including objects with large excesses of light neutron-capture
elements relative to heavy ones, originated from the r-process,
although the r-process fraction of these light neutron-capture
elements is still quite uncertain.\footnote{Indeed, the decomposition
of solar-system abundances by \citet{burris00} indicates a larger fraction
of the r-process component for Y, resulting in [Y/Zr]$_{\rm r-ss}$ =0.17,
much higher than found in r-process element-enhanced stars
\citep[see][]{hill02}.} The predictions of this abundance ratio by
existing models of the r-process exhibit rather large variations
\citep{woosley94,wanajo02,wanajo03}, presumably reflecting the
uncertainties of nuclear data and wide range of parameters of models
such as the electron fraction. The small scatter in the Y/Zr ratios
found in the VMP stars suggests the existence of some
mechanisms that regulate this abundance ratio. The abundance ratio of
Y/Zr in these stars and its small scatter could be strong constraints
on the modeling of the r-process nucleosynthesis.

%(COMPARISON WITH PREDICTIONS BY R-PROCESS MODEL HERE?)

The upper panel of Figure~\ref{fig:sryzr} shows [Sr/Y] as a function of [Ba/H].
The bulk of stars have $0.0 \le {\rm [Sr/Y]} \le +0.50$, but the scatter is much
larger than that observed in [Y/Zr]. This may reflect the large errors in Sr
abundance measurements caused by the limited number of lines used for the
measurements and the difficulty in the analysis of strong resonance lines.
Given such relatively large uncertainties, we can only claim that
the [Sr/Y] value is constant within a range of $\sim$0.3~dex. \citet{johnson02b}
also investigated the [Sr/Y] ratios in metal-poor stars, and suggested a
constant value of [Sr/Y] with a relatively large scatter.

%Indeed, there seems to exist a systematic difference between the results of the
%present work and those of Paper II, which may be due to differences in the
%treatment of the line wings and/or the processes to determine the microturbulent
%velocities.

In Figure~\ref{fig:baeu}, we show for completeness the [Ba/Eu] ratios, which
have been studied in detail by previous studies \citep[e.g., ][]{mcwilliam98}.
The average of the [Ba/Eu] values is $\sim -0.6$, agreeing with the ratio of the
r-process component in solar-system material. This indicates again that the
neutron-capture elements, at least the heavy ones, in these metal-poor stars are
dominated by the products of the r-process.

\subsection{Ba-Enhanced stars}\label{sec:yzreu}

In this section we investigate the correlation between the abundances of light
and heavy neutron-capture elements, adopting Y, Zr and Eu abundances as
indicators, rather than the Sr and Ba abundances. The first two elements represent
the light neutron-capture elements, while Eu represents the heavy neutron-capture
elements. Since the absorption lines of \ion{Y}{2}, \ion{Zr}{2}, and \ion{Eu}{2}
are much weaker than the resonance lines of \ion{Sr}{2} and \ion{Ba}{2}, the
abundances of Y, Zr, and Eu are only determined for stars having relatively high
abundances of neutron-capture elements. However, the uncertainties of abundance
measurements for these three elements are smaller than those for Sr and Ba in
general. Therefore, Y, Zr, and Eu are good probes to investigate the light and
heavy neutron-capture elements in neutron-capture-element-rich stars.

The upper and lower panels of Figure~\ref{fig:yzreu} show the
abundances of Y and Zr, respectively, as functions of the Eu
abundance. The typical abundance ratio of $N_{\rm Ba}/N_{\rm Eu}$ in
these stars is $\sim 10$, corresponding to [Ba/Eu]$\sim -0.7$. Hence,
the distribution of Eu abundance from $\log \epsilon$(Eu) = $-3$ to
$-1$ corresponds to that of the Ba abundance from $\log \epsilon$(Ba)
= $-2$ to 0 in Figure~\ref{fig:srba}.  The Y and Zr abundances show
clear correlations with the Eu abundance, in particular in the range
$\log \epsilon$(Eu)$>-2$. The scatter seen in these diagrams is much
smaller than that in the Sr--Ba diagram (Figure~\ref{fig:srba}), even
if we limit the discussion to stars with high Ba abundances. The tight
correlation between the Y (Zr) and Eu abundances suggests that the
scatter of Sr (and Ba) abundances found in the stars with high Ba
abundances in the Sr--Ba diagram is, at least partially, caused by
measurement errors in Sr and/or Ba. It should be noted that, even if
errors of abundance measurements are included, the very large
($>2$~dex) scatter in the Sr abundances among stars with low Ba
abundances remains.

The slope of the correlation found in the Y--Eu and Zr--Eu diagrams is
very interesting. Since these are diagrams of logarithmic abundances,
the increase of both elements with a fixed ratio (on a linear scale)
results in a line with a slope of unity, and the change of the ratio
appears only as a parallel shift of the line.  Since stars with
extremely low metallicity ([Fe/H]$<-2.5$) are treated here, their
abundances of neutron-capture elements are expected to have been
determined by a quite limited number of processes
\citep{audouze95}. The large scatter in the abundances ($\log
\epsilon$ values) of neutron-capture elements is interpreted as a
result of variation in the amount of yields by individual supernovae,
and subsequent dilution by interstellar matter. If the interstellar
matter contains almost no neutron-capture elements, the dilution does
not change the abundance ratios between neutron-capture elements,
while that changes their total abundances (relative to hydrogen). If
we assume fixed abundance ratios between light and heavy
neutron-capture elements, the slope found in Figure~\ref{fig:yzreu}
should be unity. However, a correlation with a slope of $\sim 1/2$ is
found in the figure. The line with a slope of 1/2 is formed by the
increase of abundances with a relation of $y\propto x^{1/2}$ (e.g.,
$N_{\rm Y} \propto N_{\rm Eu}^{1/2}$). Such an behavior is quite
unexpected, if the abundances of neutron-capture elements relative to
hydrogen are determined by the yields from explosive processes like
supernovae and the dilution by interstellar matter that contains no
neutron-capture elements.  Moreover, the scatter is larger at lower Eu
abundances.

%The slope of the correlation found in the Y--Eu and Zr--Eu diagrams is
%very interesting. Since these are diagrams of logarithmic abundances,
%the increase of both elements with a fixed ratio (on a linear scale)
%results in a line with a slope of unity, and the change of the ratio
%appears only as a parallel shift of the line. The line with a slope of
%1/2 is formed by the increase of abundances with a relation of
%$y\propto x^{1/2}$ (e.g., $N_{\rm Y} \propto N_{\rm Eu}^{1/2}$). Such
%an behavior is quite unexpected, if the abundances of neutron-capture
%elements relative to hydrogen are determined by the yields from
%explosive processes like supernovae and the dilution by interstellar
%matter.  Moreover, the scatter is larger at lower Eu abundances.

How might we explain the observed distribution of Y and Eu
abundances? If we assume two processes, which produce different
ratios of Y(Zr)/Eu, some insight can be obtained. The solid lines in
the upper panel of Figure~\ref{fig:yzreu} indicate the increase of Eu
and Y abundances with a fixed ratio of $\delta N_{\rm Y}/\delta N_{\rm
Eu} = 5$, assuming two different initial values ($\log \epsilon$(Y)$=
-0.5$ and $-2.8$ at $\log \epsilon$(Eu)$=-3.5$). Most stars in the
diagram can be explained by changing the initial Y abundances between
these two values. The dotted lines correspond to the case of $\delta
N_{\rm Y}/\delta N_{\rm Eu} = 2$ and initial values of $\log
\epsilon$(Y)$= -0.2$ and $-3.2$ at $\log \epsilon$(Eu)$=-3.5$. The large
cross in the figure indicates the values of the r-process
component for these elements in the Solar System \citep{arlandini99},
which can also be explained by the increase of Y and Eu in such a
manner, although it must be kept in mind that the r-process
contribution to solar-system material for light neutron-capture
elements is still somewhat uncertain. 

One might intuit that the increase of Y abundances with respect to Eu
with a fixed ratio corresponds to enrichment by the main r-process,
while the initial values assumed above are determined by the process
that produced light neutron-capture elements with little accompanying
heavy species. If this is true, the neutron-capture elements in stars
near the line with a slope of unity are provided by the main
r-process, while all other stars are more or less affected by the
process that produces light neutron-capture elements.  The above
comparisons of solid and dotted lines with data points indicate that
the main r-process produces the Y/Eu abundance ratio of $\delta N_{\rm
Y}/\delta N_{\rm Eu} = 2\sim 5$, while another process yields these
elements with a quite large distribution ($\sim 2$~dex) in the ratio.

The solid and dotted lines in the lower panel of
Figure~\ref{fig:yzreu} indicates the increase of Zr and Eu abundances
with a fixed ratio of $\delta N_{\rm Zr}/\delta N_{\rm Eu} = 20$ and
10, respectively, assuming two different initial values. The Zr and Eu
abundances of metal-poor stars as well as that of the r-process
component in the Solar System \citep{arlandini99}, indicated by the
cross in this figure, can also be explained by changing the initial
[Zr/Eu] ratio.

In the above discussion, the effect of the nucleosynthesis process producing
light neutron-capture elements is regarded as the source of the difference of
the ``initial values'' of Y and Zr. However, this does not necessarily mean that
these processes have operated in advance of the main r-process. The time scale
of the contribution of each process is dependent on the progenitor mass, which
is still unclear, as discussed in subsection~\ref{sec:srba}. While a fixed
abundance ratio of [Y/Eu] and [Zr/Eu] can be assumed for the main r-process,
large distributions of these abundance ratios are required for the other process
to explain the large scatter in the Y and Zr abundances at low Eu abundance.
Further observational studies to determine the elemental abundance patterns
produced by this process, covering a wider atomic number range, are clearly
required.

\section{Summary and Concluding Remarks}

We have measured elemental abundances for 18 very metal-poor stars using
spectra obtained with the Subaru Telescope High Dispersion
Spectrograph. The metallicity range covered by our sample is $-3.1
\lesssim$ [Fe/H] $\lesssim -2.4$. While the abundances of carbon,
$\alpha$-elements, and iron-peak elements determined for these stars
show similar trends to those found by previous work, we found an
exceptional star, BS~16934--002, a giant with [Fe/H] = $-2.8$ having
large over-abundances of Mg, Al and Sc. Further abundance studies for
this object are strongly desired.

By combining our new results with those of previous studies, we
investigated the distribution of neutron-capture elements in very
metal-poor stars, focusing on the production of the light
neutron-capture elements (Sr, Y, and Zr), and found the following
observational results:

\noindent
(1) A large scatter is found in the abundance ratios between the light
and heavy neutron-capture elements (e.g., Sr/Ba) for stars with
low abundances of heavy neutron-capture elements. Most of these stars
have extremely low metallicity ([Fe/H]$\lesssim -3$). 

\noindent
(2) Stars with high abundances of heavy neutron-capture elements appear
in the metallicity range of [Fe/H]$\gtrsim -3$. The observed scatter
in the ratios between light and heavy neutron-capture elements is much
smaller in these stars. In particular, the Y and Zr abundances exhibit
a clear correlation with Eu abundance in stars with high Eu
abundances, but the trend is not explained by the increases of
light and heavy elements with fixed abundance ratios.

\noindent
(3) The Y/Zr ratio is similar in stars with high and low abundances of
heavy neutron-capture elements. The values of the [Y/Zr] indicate
these are not products of the main nor weak components of the s-process, but
must have an origin related to the r-process. 

These observational results indicate the existence of the process that
yielded light neutron-capture elements (Sr, Y, and Zr) with little
contribution to heavy ones (e.g., Ba, Eu). This process seems to be
different from the weak s-process. Such a process has been suggested by
previous studies \citep[e.g., ][]{truran02,johnson02b,travaglio04}, as
mentioned in Section~\ref{sec:intro}. The above results suggest that
this process appears even in extremely metal-poor stars
([Fe/H]$\sim -3.5$), but is seen as well in stars with higher metallicity
and higher abundances of heavy neutron-capture elements. The ratios of
light to heavy neutron-capture elements (e.g., Sr/Ba, Y/Eu) formed by
this process have a wide distribution, while the abundance ratios of
elements among light neutron-capture elements (e.g., Y/Zr), as well as
those among heavy ones (e.g., Ba/Eu), are almost constant.

These observational results provide new constraints on modeling r-process
nucleosynthesis, and identifying its likely astrophysical sites. However,
further observational studies are required. In particular, measurements for
lower metallicity stars ([Fe/H]$<-3.5$) are very important to understand the
process that produced light neutron-capture elements in the very early Galaxy.
More detailed abundance studies for stars showing large excesses of light
neutron-capture elements, with low abundances of heavier ones, will provide
definitive constraints on the modeling of that process.

\acknowledgments

T.C.B. acknowledges partial support from a series of grants awarded by the US
National Science Foundation, most recently, AST 04-06784, as well as from grant
PHY 02-16783; Physics Frontier Center/Joint Institute for Nuclear Astrophysics
(JINA).

\clearpage

\input{tables.tex}
\clearpage
\input{figcap.tex}

\end{document}

%% file: tables.tex
\begin{deluxetable}{rlcccccc}
\tablewidth{0pt}
\tablecaption{OBJECT LIST AND OBSERVING LOG \label{tab:obs}}
\startdata
\tableline
\tableline
ID & Object name & RA(2000) & Dec(2000) & Exposure  & S/N\tablenotemark{a} & Obs. Date (JD) &  Radial velocity  \\
            &          &            &        &  (min)     &             & (km s$^{-1}$)    \\
\tableline
 1 & HD122563      & 14 02 31.8 & +09 41 10 & 15 &1500 &  2 Feb. 2002 & $ -25.81 \pm 0.17$ \\
 2 &BS~16033--008 & 13 05 56.1 & +25 10 17 & 60 & 100 & 16 Apr. 2001 & $ -30.43 \pm 0.27$ \\
 3 &BS~16080--054 & 16 39 18.5 & +60 57 42 & 20 &  89 & 23 July 2001 & $-120.80 \pm 0.25$ \\
 4 &BS~16083--172 & 14 50 41.6 & +48 45 14 & 40 &  86 & 31 July 2001 & $-165.41 \pm 0.33$ \\ 
 5 &BS~16084--160 & 16 28 50.7 & +54 37 03 & 40 & 102 & 23 July 2001 & $-118.71 \pm 0.24$ \\
 6 &BS~16089--013 & 13 49 22.1 & +35 44 53 & 40 &  86 & 31 July 2001 & $-257.18 \pm 0.31$ \\
\enddata
\tablenotetext{a}{ S/N ratio per resolution element (6~km~s$^{-1}$) estimated from the photon counts at 4000~{\AA} (see text)}
\end{deluxetable}

\begin{deluxetable}{lcccccccclcc}
\rotate
\tablewidth{0pt}
\tablecaption{PHOTOMETRIC DATA \label{tab:photometry}}
\startdata
\tableline
\tableline
Object name & $B$ & $V$ & $R$ & $I$ & $J$ & $H$ & $K$ & $E(B-V)$ & $T_{\rm eff}(V-K)_{0}$ & $\Delta(T_{\rm eff})$\tablenotemark{a} \\  
\tableline
HD~122563 \tablenotemark{b} &   7.11  &   6.20  &   5.65   &  5.11   &   4.79  &   4.03  &   3.73  &    0.00 & 4605 &$-$11 \\
BS~16033--008 &  14.416 &  13.786 &  13.367 &  12.943 &  12.327 &  11.881 &  11.831 &    0.01 & 5240 &$-$14 \\
BS~16080--054 &  13.550 &  12.776 &  12.285 &  11.786 &  11.082 &  10.604 &  10.504 &    0.00 & 4811 &$-$5 \\
BS~16083--172 &  14.058 &  13.434 &  13.025 &  12.594 &  11.991 &  11.578 &  11.496 &    0.01 & 5264 & 4 \\
BS~16084--160 &  14.002 &  13.144 &  12.630 &  12.103 &  11.358 &  10.849 &  10.740 &    0.00 & 4670 &$-$33 \\
BS~16089--013 &  14.065 &  13.314 &  12.842 &  12.333 &  11.677 &  11.208 &  11.119 &    0.00 & 4898 & 62 \\
BS~16543--097 &  13.261 &  12.570 &  12.113 &  11.657 &  11.009 &  10.542 &  10.463 &    0.01 & 5038 &$-$19 \\
\enddata
\tablenotetext{a}{$\Delta(T_{\rm eff}) = T_{\rm eff}(V-K)_{0} - <T_{\rm eff}(B-V)_{0}, T_{\rm eff}(V-R)_{0}, T_{\rm eff}(V-I)_{0}, T_{\rm eff}(R-I)_{0}>$}

\tablenotetext{b}{$R$ and $I$ data are adopted from \citet{bessell84}, while others are from \citet{honda04b}.}
\end{deluxetable}%

%% file: figcap.tex
\begin{figure} 
\includegraphics[width=8.5cm]{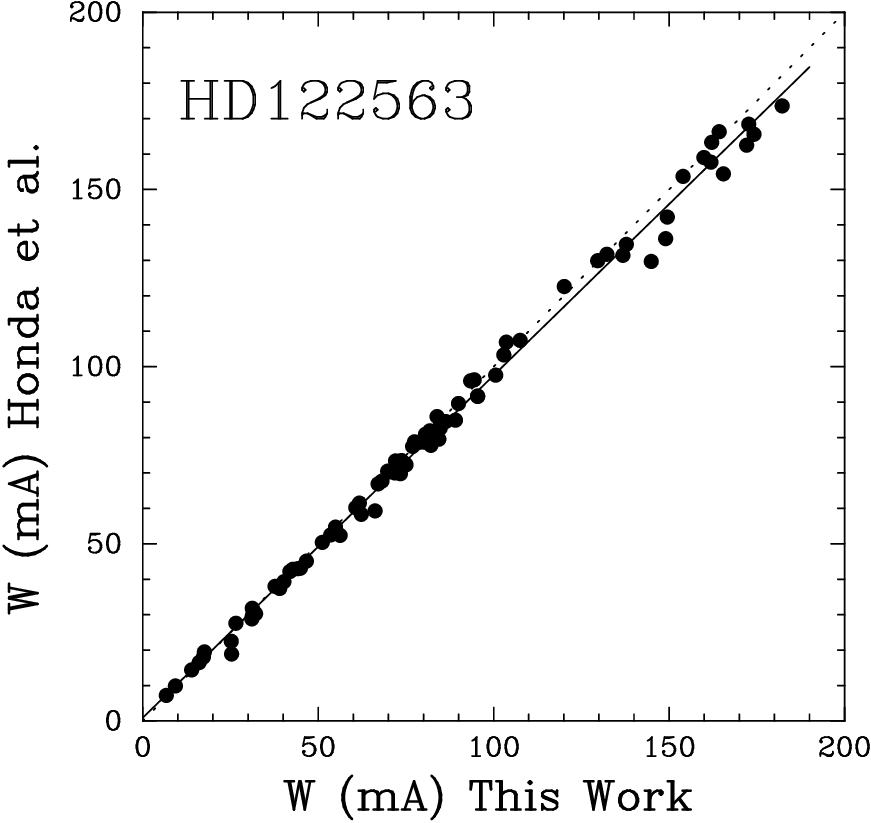}
\includegraphics[width=8.5cm]{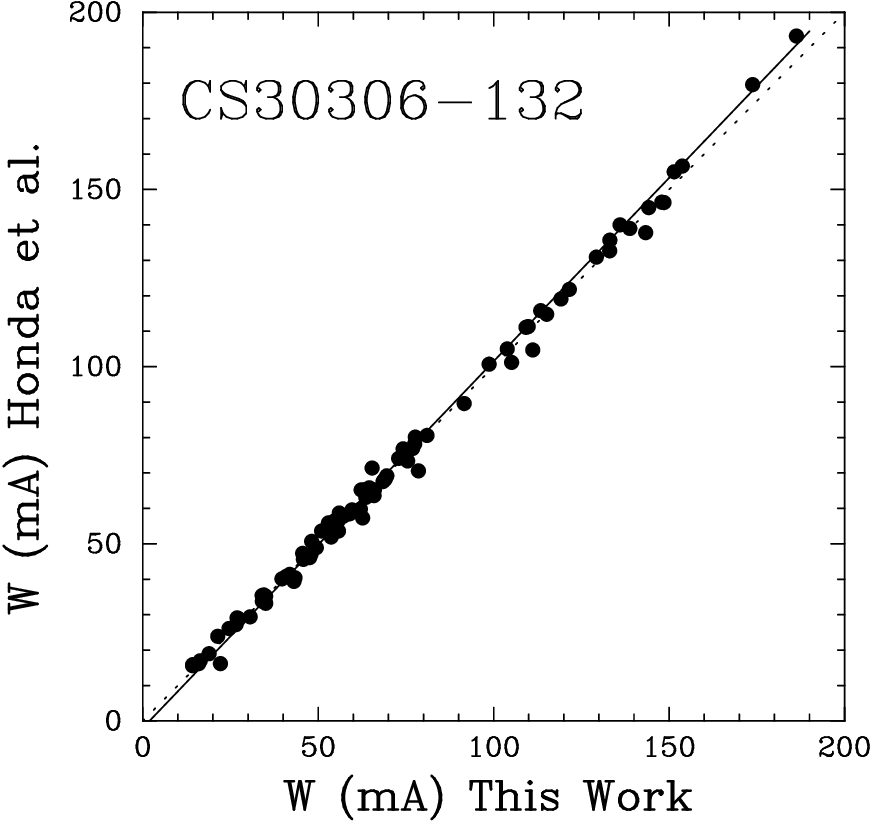} 

\caption[]{Comparisons of the equivalent widths of \ion{Fe}{1} lines
measured by Paper I and by the present work for HD~122563 and
CS~30306--132. The solid line indicates the result of a least square
fit, while the dotted line shows a line with a slope of unity.}

\label{fig:ew1} 
\end{figure} 

\begin{figure}
\includegraphics[width=8.5cm]{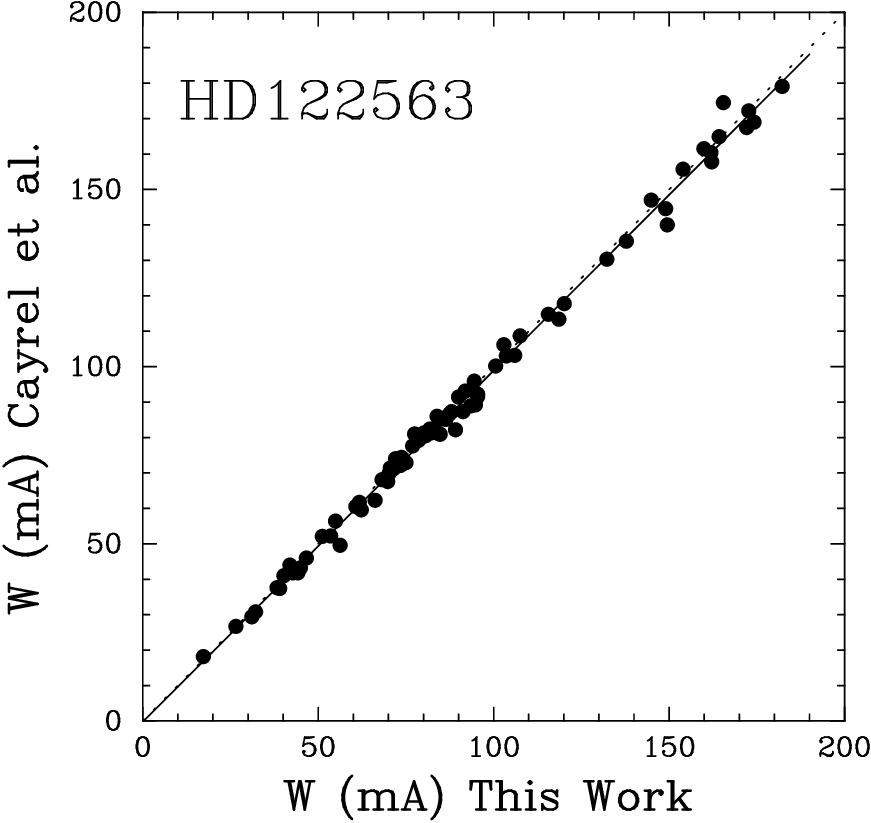}
\includegraphics[width=8.5cm]{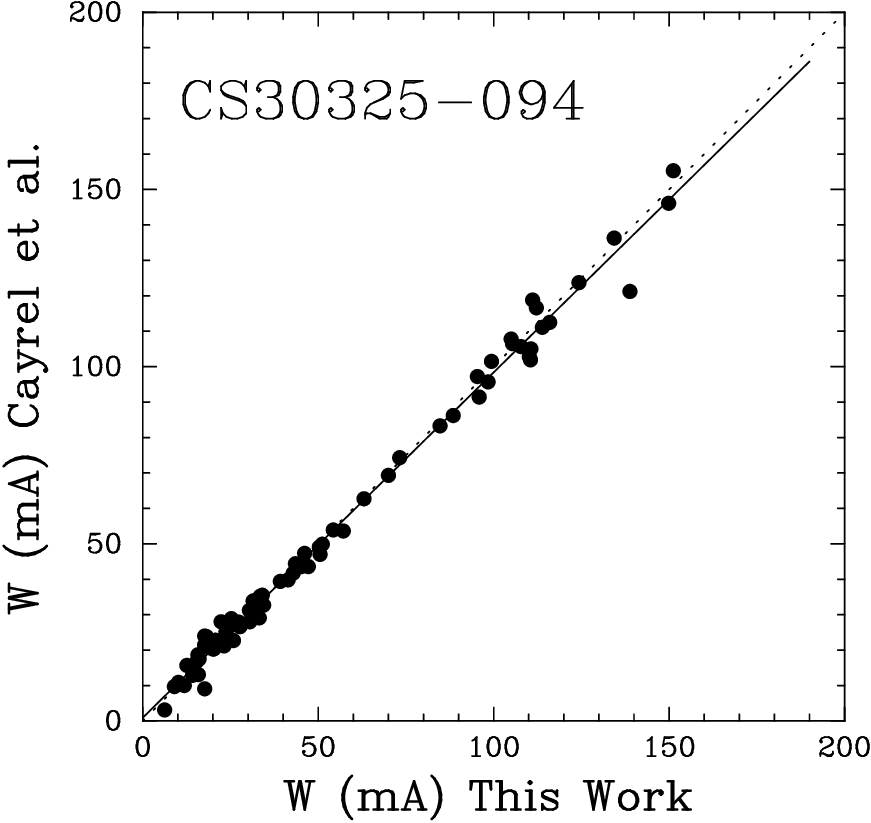}

\caption[]{Comparisons of the equivalent widths of \ion{Fe}{1}
measured by \citet{cayrel04} and by the present work for HD~122563 and
CS~30325--094. The meanings of the lines are the same as those in
Figure~\ref{fig:ew1}.}  

\label{fig:ew2} 
\end{figure}

\begin{figure}
\includegraphics[width=8.5cm]{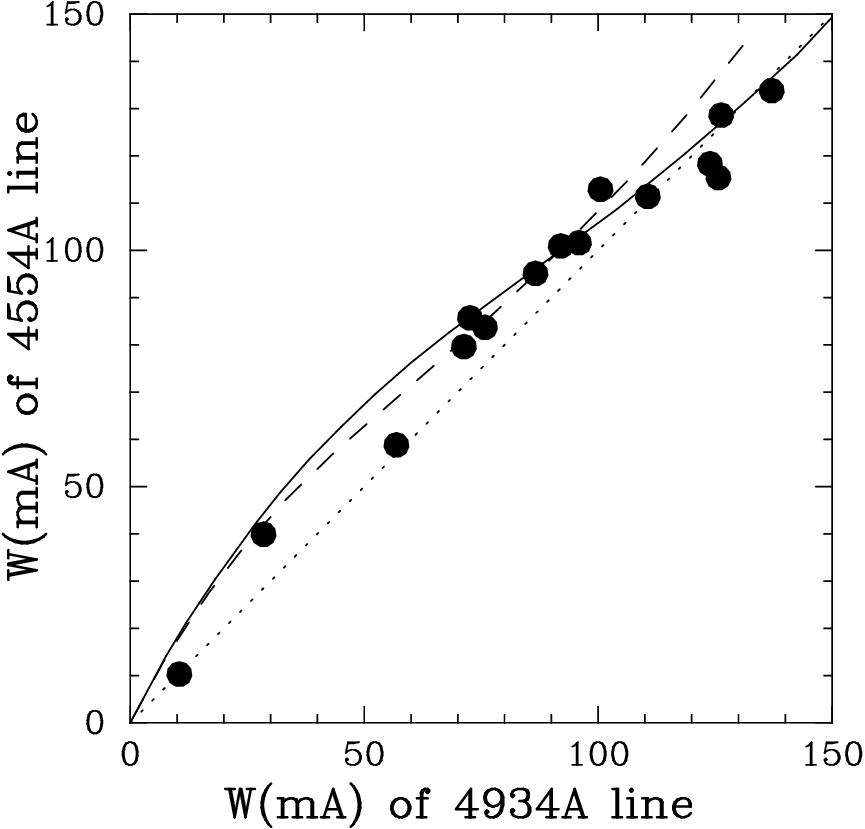}

\caption[]{Comparisons of the equivalent widths of \ion{Ba}{2}
4554~{\AA} and 4934~{\AA} lines (dots). The solid line indicates the
comparisons of calculated equivalent widths including hyperfine
splitting for the model atmosphere with $T_{\rm eff}=$ 4900~K, $\log
g=$ 1.7, and [Fe/H] = $-2.8$, assuming $v_{\rm turb}=$
1.9~km~s$^{-1}$. The dashed line shows the same comparison, but for
the calculations not including hyperfine splitting.  In the range
of weak lines ($W\lesssim 50$~m{\AA}), the 4554~{\AA} line is
predicted to be stronger than the 4934~{\AA} line. The difference
becomes smaller in the stronger lines due to saturation effects. In
the range of strongest lines ($W\gtrsim 120$~m{\AA}), the calculations
including hyperfine splitting predict almost equal values of
equivalent widths for both lines, while the calculations not
including hyperfine splitting predict larger equivalent widths for the
4554~{\AA} line than for the other line. The behavior of of measured
equivalent widths for the strongest lines can be well explained by the
calculations including hyperfine splitting.}

\label{fig:ewba} 
\end{figure}

\begin{figure} 
\includegraphics[width=8.5cm]{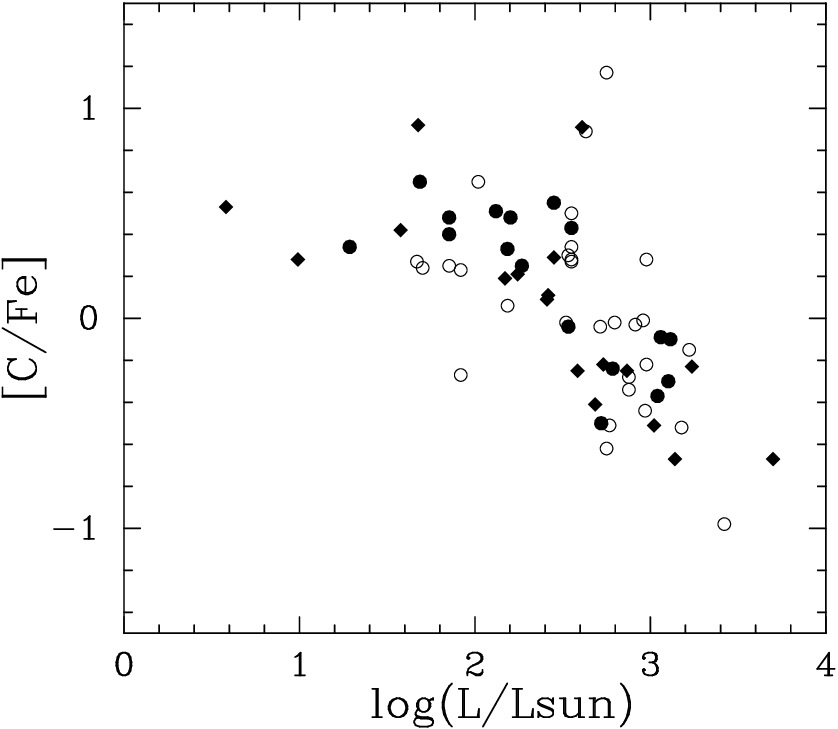} 

\caption[]{[C/Fe] as a function of luminosity estimated from $T_{\rm
eff}$ and $\log g$ (see text).}

\label{fig:cl} 
\end{figure}

\begin{figure} 
\includegraphics[width=8.5cm]{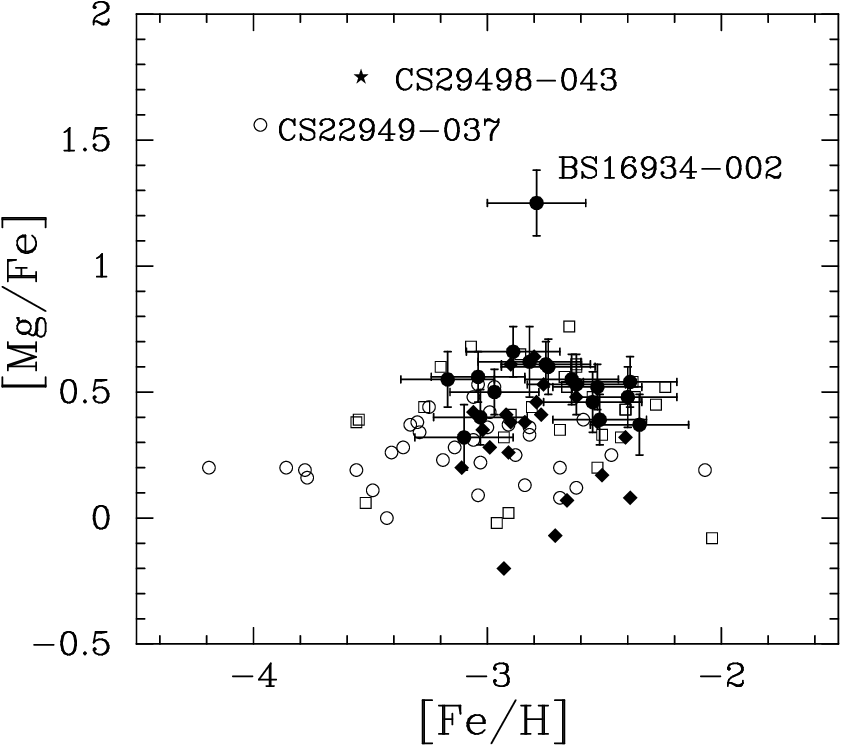} 

\caption[]{[Mg/Fe] as a function of [Fe/H]. The results of the present
study are shown by the filled circles with error bars. Results of
previous studies are also shown by filled diamonds (Paper II), open
circles \citep{cayrel04,depagne02}, open squares
\citep{cohen04,carretta02}, and a filled star \citep{aoki04}. The three Mg
enhanced stars (see text) are identified in the panel.}

\label{fig:mgfe} 
\end{figure}

\begin{figure} 
\includegraphics[width=8.5cm]{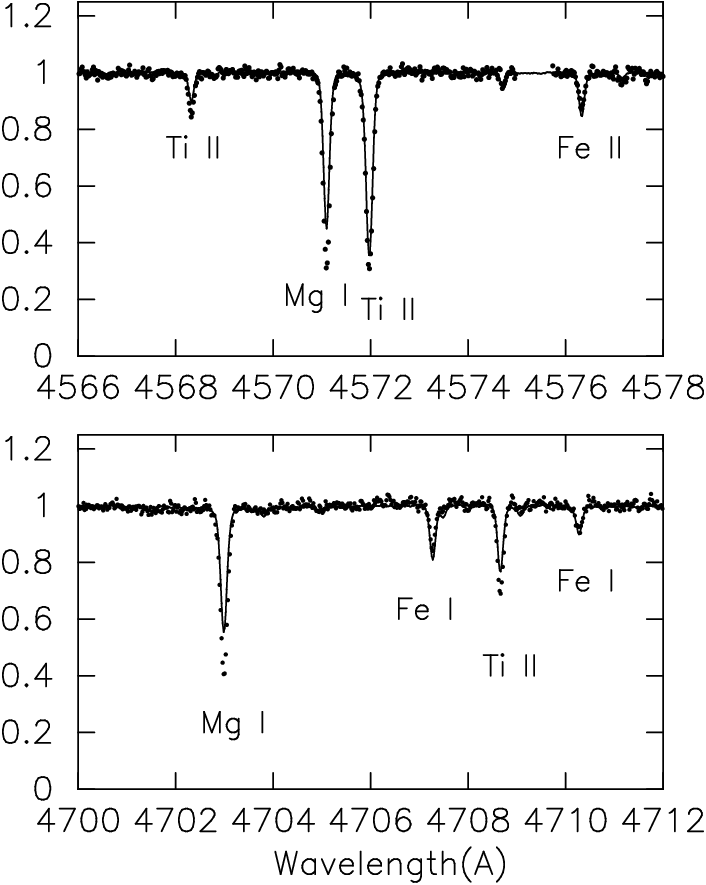}

\caption[]{Comparisons of the spectrum of BS~16934--002 (filled
circles) with that of HD~122563 (the line). These two stars have quite
similar atmospheric parameters. The strengths of
\ion{Fe}{1} and \ion{Fe}{2} lines are similar between the two stars,
while the \ion{Mg}{1} 4571 and 4703~{\AA} lines of BS~16934--002 are
much stronger than those of HD~122563. A small portion of the spectrum at
around 4575~{\AA} of BS~16934--002 was lost because of a defect of the
detector.}

\label{fig:sp_mg}
\end{figure}

\begin{figure} 
\includegraphics[width=8.5cm]{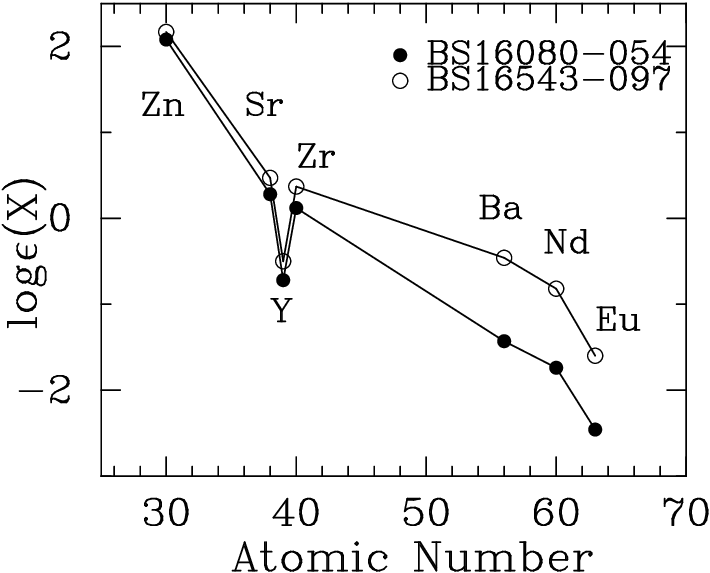}

\caption[]{Elemental abundance patterns of BS~16080--054 (filled
circles) and BS~16543--097 (open circles) from Zn to Eu. The
abundances of heavy neutron-capture elements are significantly different
between the two stars, while metallicity and abundances of light
neutron-capture elements are quite similar.}

\label{fig:abpat1} 
\end{figure}

\begin{figure} 
\includegraphics[width=8.5cm]{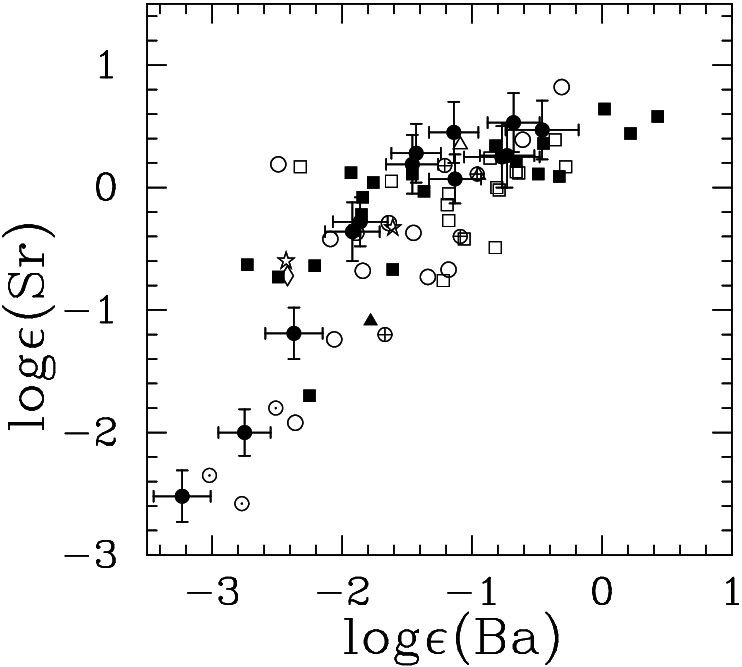}

\caption[]{Sr abundances as a function of the Ba abundance for very
metal-poor stars ([Fe/H]$\leq-2.5$). Stars having significant excesses
of carbon and s-process elements are excluded. The present results are
shown by filled circles with error bars. Results of previous studies
are also shown by filled squares (Paper II), open circles
\citep{mcwilliam95,mcwilliam98}, open squares
\citep{cohen04,carretta02}, circles with cross \citep{johnson02b}, open
triangles \citep{burris00}, open stars \citep{ishimaru04}, circles
with enclosed point \citep{francois03}, a open diamond
\citep{depagne02}, and a filled triangle \citep{aoki04}.}

\label{fig:srba} 
\end{figure}

\begin{figure} 
\includegraphics[width=6.5cm]{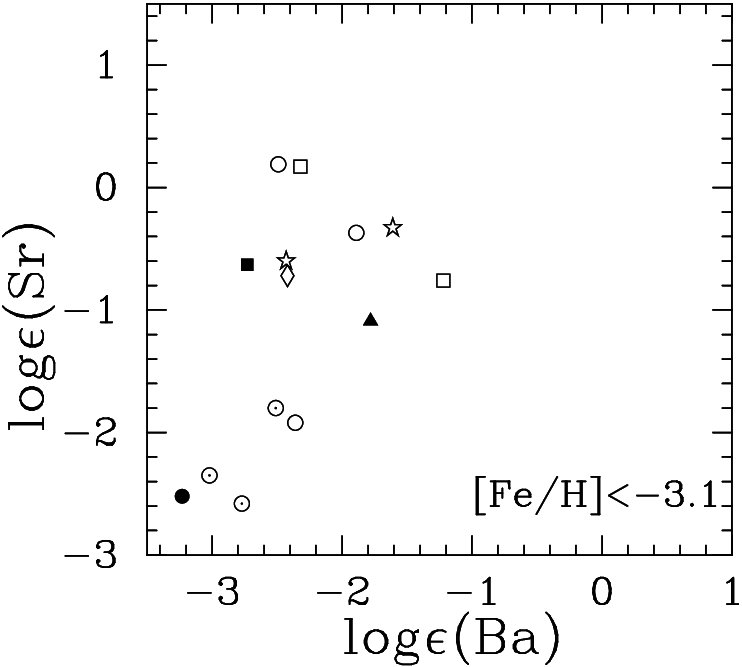} \\
\includegraphics[width=6.5cm]{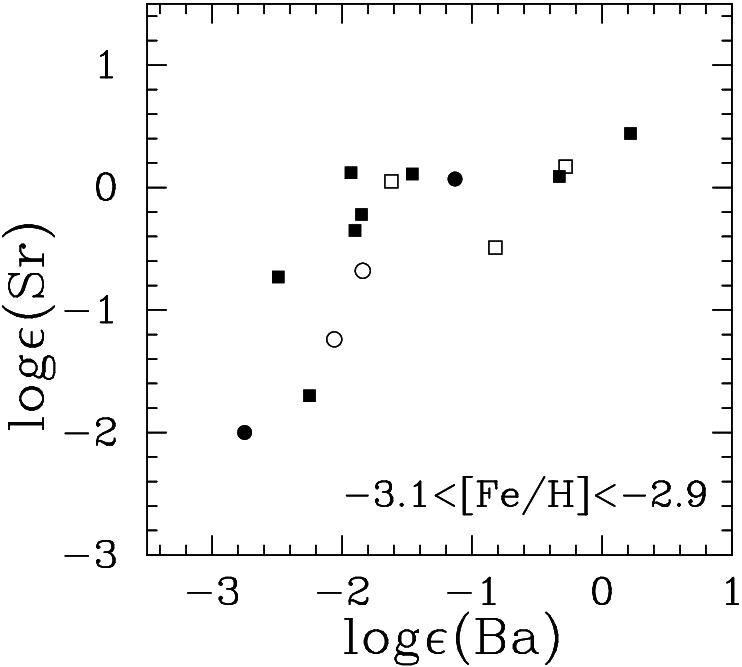} \\
\includegraphics[width=6.5cm]{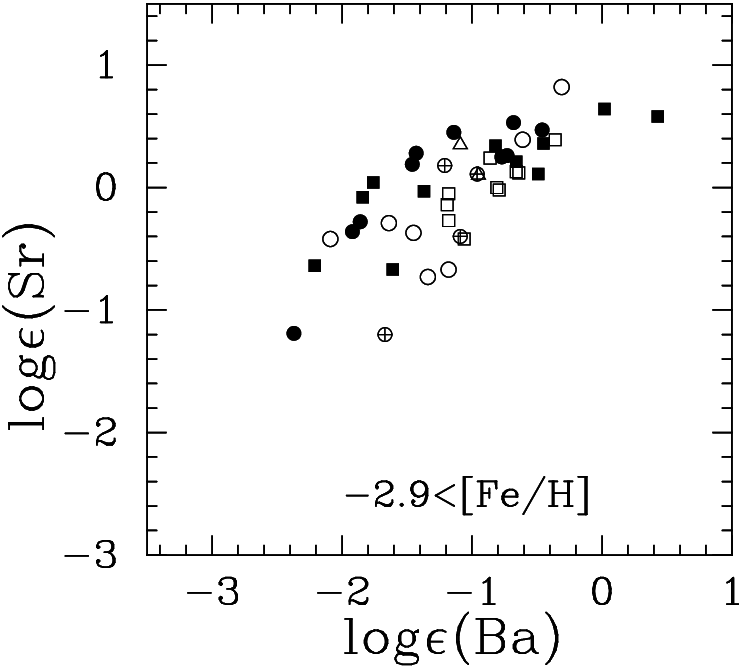} 

\caption[]{The same as Fig.~\ref{fig:srba}, but for stars with
[Fe/H]$\leq -3.1$ (top), $-3.1<$[Fe/H]$\leq -2.9$ (middle), and
$-2.9<$[Fe/H]$\leq -2.5$ (bottom). The open diamond and the filled
triangle in the top panel denote the abundances of CS~22949--037
and CS~29498--043, respectively, which have very low iron abundances
but large enhancements of carbon and the $\alpha$--elements.}

\label{fig:srba3} 
\end{figure}

\begin{figure} 
\includegraphics[width=8.5cm]{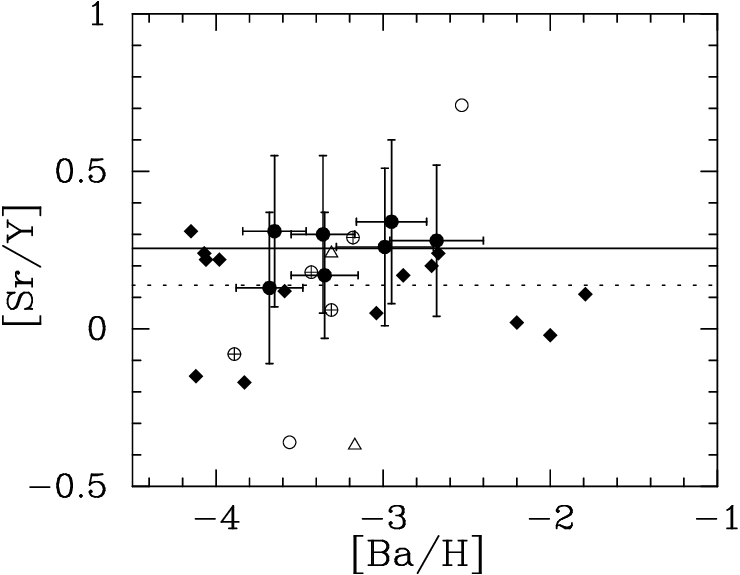} \\
\includegraphics[width=8.5cm]{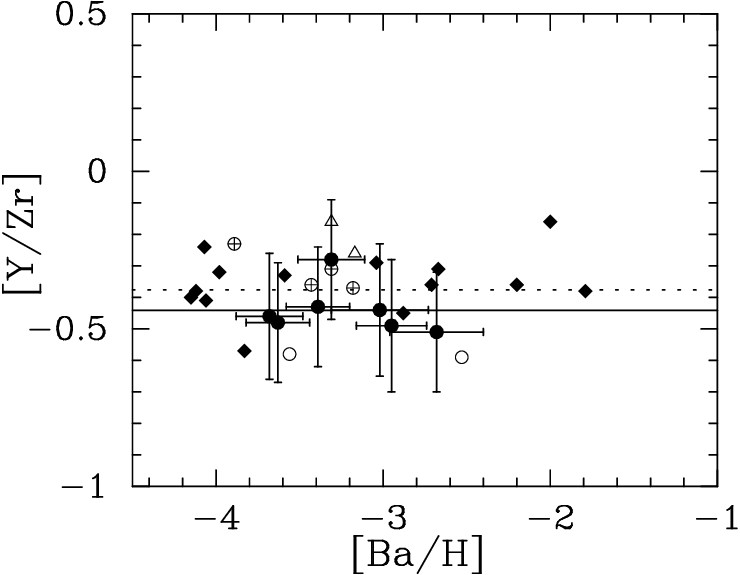}

\caption[]{[Sr/Y] (top) and [Y/Zr] (bottom) as functions of
[Ba/H]. The symbols have the same meanings as in
Figure~\ref{fig:srba}. The solid line shows the average value of
[Sr/Y] or [Y/Zr] of 7 stars in our sample in which all of these three
elements are detected. The dotted line indicates the average value of
all stars shown in the figure (29 stars) including objects studied by
previous work.}

\label{fig:sryzr} 
\end{figure}

\begin{figure} 
\includegraphics[width=8.5cm]{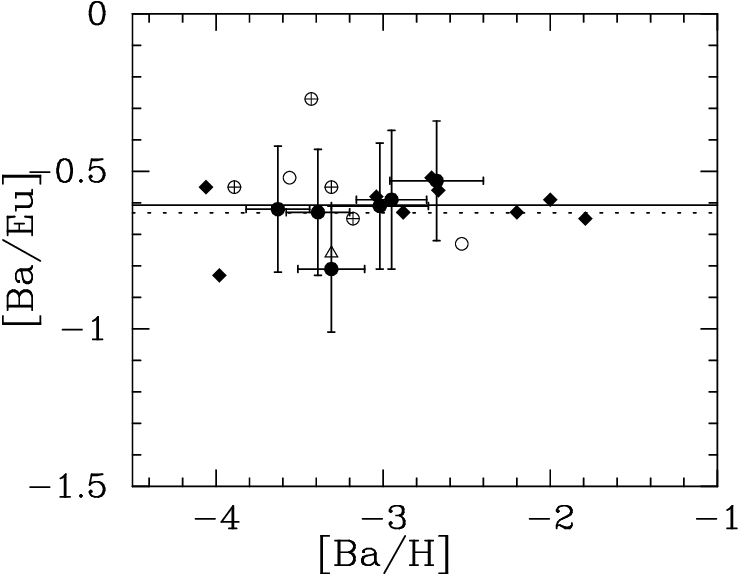} \\
\caption[]{The same as Fig.~\ref{fig:sryzr}, but for [Ba/Eu]}
\label{fig:baeu} 
\end{figure}

\begin{figure} 
\includegraphics[width=8.5cm]{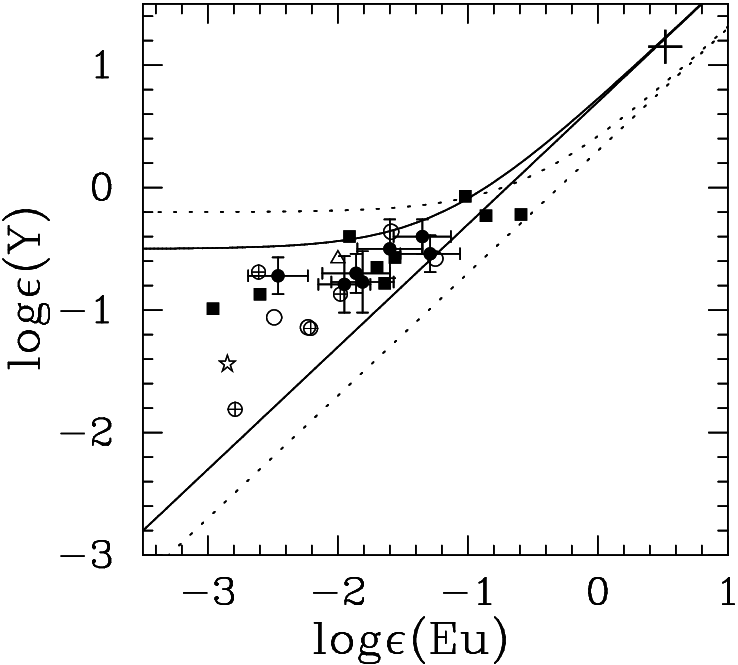} \\
\includegraphics[width=8.5cm]{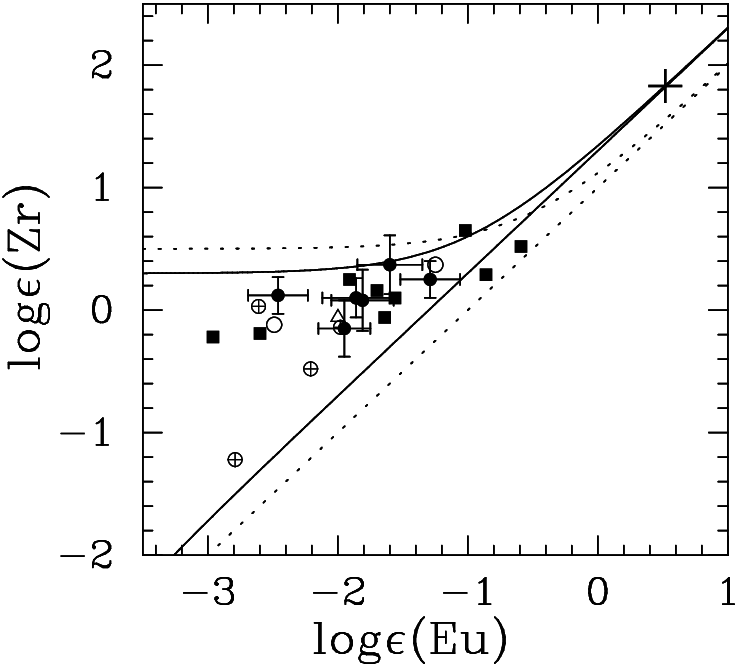}

\caption[]{Abundances of Y and Zr as functions of the Eu
abundance. The meaning of the symbols is the same as in
Fig.~\ref{fig:srba}. The solid and dotted lines show the enrichment
of related elements assuming different initial abundances and a
constant Y/Eu or Zr/Eu ratios in the yields of the main r-process. See
text for details. The cross indicates the abundances of the r-process
component of solar-system material estimated by \citet{arlandini99}.}

\label{fig:yzreu} 
\end{figure}